\def\year{2022}\relax
\documentclass[letterpaper]{article} 
\usepackage{aaai22}  
\usepackage{times}  
\usepackage{helvet}  
\usepackage{courier}  
\usepackage[hyphens]{url}  
\usepackage{graphicx} 
\usepackage{natbib}  
\usepackage{caption} 
\DeclareCaptionStyle{ruled}{labelfont=normalfont,labelsep=colon,strut=off} 
\usepackage{algorithm}
\usepackage{algorithmic}
\usepackage{booktabs}
\usepackage{xcolor}

\usepackage{amssymb}
\usepackage{amsmath}
\usepackage{latexsym}

\usepackage{lineno}
\usepackage{csquotes}
\usepackage{pdflscape}

\usepackage{subcaption}

\usepackage{newfloat}
\usepackage{listings}
\floatstyle{ruled}
\newfloat{listing}{tb}{lst}{}
\floatname{listing}{Listing}
\title{Twitter Corpus of the \#BlackLivesMatter Movement and Counter Protests:\\ 2013 to 2021}
\author {
    Salvatore Giorgi,\textsuperscript{\rm 1,2}
    Sharath Chandra Guntuku,\textsuperscript{\rm 2}
    McKenzie Himelein-Wachowiak,\textsuperscript{\rm 1} \\
    Amy Kwarteng,\textsuperscript{\rm 1}
    Sy Hwang,\textsuperscript{\rm 2}
    Muhammad Rahman,\textsuperscript{\rm 1}
    Brenda Curtis\textsuperscript{\rm 1}\\
}
\affiliations {
    \textsuperscript{\rm 1} National Institute on Drug Abuse\\ \textsuperscript{\rm 2} University of Pennsylvania \\
    sal.giorgi@nih.gov, brenda.curtis@nih.gov
}

\begin{document}

\maketitle

\begin{abstract}
Black Lives Matter (BLM) is a decentralized social movement protesting violence against Black individuals and communities, with a focus on police brutality. 
The movement gained significant attention following the killings of Ahmaud Arbery, Breonna Taylor, and George Floyd in 2020. 
The \#BlackLivesMatter social media hashtag has come to represent the grassroots movement, with similar hashtags counter protesting the BLM movement, such as \#AllLivesMatter, and \#BlueLivesMatter. 
We introduce a data set of 63.9 million tweets from 13.0 million users from over 100 countries which contain one of the following keywords: \emph{BlackLivesMatter}, \emph{AllLivesMatter}, and \emph{BlueLivesMatter}. 
This data set contains all currently available tweets from the beginning of the BLM movement in 2013 to 2021. 
We summarize the data set and show temporal trends in use of both the \emph{BlackLivesMatter} keyword and keywords associated with counter movements. 
Additionally, for each keyword, we create and release a set of Latent Dirichlet Allocation (LDA) topics (i.e., automatically clustered groups of semantically co-occuring words) to aid researchers in identifying linguistic patterns across the three keywords. 
\end{abstract}

\section{Introduction}
\label{sec:intro}

The murder of George Floyd, an unarmed Black man, at the hands of police started a wave of global protests across the second half of 2020. In the U.S., the number of locations holding protests related to this event, as well as other killings of unarmed Black individuals such as Breonna Taylor and Ahmaud Arbery, outnumbered any other demonstration in U.S. history~\cite{putnam2020floyd}. Notably, demonstrations were not limited to larger, urban areas, with protests occurring in all 50 states. 
An overwhelming number of these events were associated with the Black Lives Matter (BLM) movement~\cite{kishi2020demonstrations}, a decentralized grass roots movement protesting police brutality and violence against Black individuals.
The global response to George Floyd's murder was in part due to the lose network of BLM related organizations, as well as previous demonstrations dating back to the movement's origins following the killing of unarmed Black teenager Trayvon Martin and the subsequent acquittal of perpetrator George Zimmerman.
While support for BLM has fluctuated since its inception, support during the summer of 2020 had increased across all ethnic and racial groups~\cite{horowitz2020amid} and increased attention towards other Black victims of police violence~\cite{wu2021say}. 

Central to the BLM movement is advocacy against police violence toward Black individuals, which perpetuates negative health and psychological repercussions in Black individuals and communities.
Multiple studies have shown the presence of racial bias in police violence~\cite{ross2015multi,gbd2021fatal}, with police use of force being one of the leading causes of death for Black men between 25 and 29 years of age~\cite{edwards2019risk}.
There is also evidence showing that police killings have negative effects on mental health in Black populations~\cite{bor2018police,williams2018assessing}.
Increases in depression and post-traumatic stress disorder in Black individuals are often higher than White individuals as a result of shared community violence~\cite{galovski2016exposure}.
Similar negative mental health effects in Black and Latin adolescents have been shown to be related to online exposure to traumatic events, such as widely shared videos of police killings~\cite{tynes2019race}.
The emotional impact of the murder of George Floyd was immediately felt: more than a third of the US population reported sadness and anger, with increased rates among Black Americans~\cite{eichstaedt2021emotional}. 
Furthermore, a sentiment analysis of tweets showed that May 31, 2020, six days after the death of George Floyd, was the saddest day in Twitter's history~\cite{saddest_day}.

\begin{table*}[!htb]
\centering
\resizebox{1\textwidth}{!}{
\begin{tabular}{rccccccc} \toprule
                          & Tweets     & Users      & Retweets   & Replies   & Geotagged & User Location & Top Languages      \\ \hline
\textit{All}              & 63,884,799 & 13,061,316 & 47,083,420 & 3,266,120 & 86,641 & 10,820,854 & en, fr, es, pt, ja  \\
\textit{BlackLivesMatter} & 56,693,715 & 12,322,212 & 42,693,046 & 2,590,724 & 77,257 & 9,552,502 & en, fr, es, pt, ja  \\
\textit{AllLivesMatter}   & 4,343,704  & 1,845,937  & 2,287,247  & 564,714   & 9,928 & 698,252 & en, es, nl, ja, fr  \\
\textit{BlueLivesMatter}  & 5,075,833  & 1,224,933  & 3,494,159  & 306,711   & 2,329 & 938,335 & en, fr, es, ja, de  \\ \bottomrule
\end{tabular}}
\caption{Descriptive counts for the entire data set and each keyword. Note that tweets can contain more than one keyword and can therefore be included in more than one row. ISO 639-1 Language codes: en = English, fr = French, es = Spanish, pt = Portuguese, ja = Japanese, nl = Dutch, de = German.}
\label{table: stats}
\end{table*}

Given the global reach of the BLM movement, as well as the mental and physical health impacts of violence on Black communities (a central theme of the movement), we open-source a large-scale data set to facilitate associated research in the areas of computational social science, communications, political science, natural language processing, and machine learning. 
In the past, similarly themed, though much smaller in scope, BLM data sets have been used for studying discourse in protest and counter protest movements~\cite{gallagher2018divergent,blevins2019tweeting}, predicting retweets~\cite{keib2018important}, examining the role of social media in protest movements~\cite{mundt2018scaling,ince2017social,wilkins2019whose}, examining changes in implicit and explicit racial attitudes~\cite{sawyer2018implicit}, and exploring narrative agency~\cite{yang2016narrative}. 
Research has also shown that Russian disinformation campaigns have infiltrated the BLM conversation on social media~\cite{aceves2018virtual}, in which case this data set could be used to study these campaigns. The current data set has been used for evaluating automatic event extraction systems in the context of socio-political events~\cite{giorgi-etal-2021-discovering,hurriyetoglu-etal-2021-challenges}.

These data are useful because they showcase the entire timeline of a large, ongoing social movement (Black Lives Matter) and its counter protests (All Lives Matter and Blue Lives Matter). To our knowledge, no other Twitter data sets exist that cover the entire span of the Black Lives Matter movement to date.

All researchers interested in systemic racism, social movements, grassroots campaigns, racial inequality, police brutality and counter protests, especially those working in the fields of computational social science, computational linguistics, communications, and political science, can benefit from this data.

\begin{figure*}[!htb]
\minipage{1.0\textwidth}
  \includegraphics[width=\linewidth]{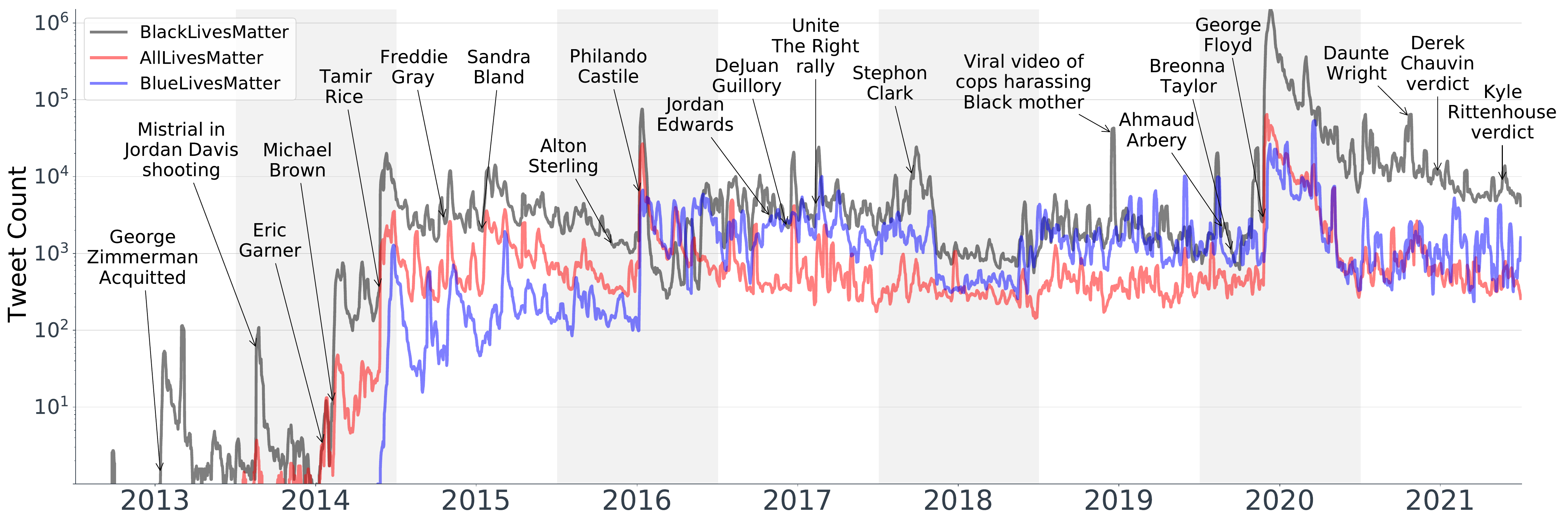}
  \caption{Seven day moving average of monthly tweet count from 2013 to 2021 of all three keywords. We include annotations for high profile events associated with the BLM movement.}\label{fig:time series}
\endminipage
\end{figure*}

\section{Data Description}
\label{sec:data description}

All data is available through Zenodo~\cite{salvatore_giorgi_2022_5835260}. 

\subsection{Tweets}

Tweets containing the keywords \emph{BlackLivesMatter}, \emph{AllLivesMatter}, and \emph{BlueLivesMatter} were collected through the Twitter API from January 2013 to December 31, 2021. 
Table \ref{table: stats} contains counts of total number of tweets and users for the entire data set and each keyword. 
It also includes counts for the following: retweets (original tweets which are shared by other users on the platform), replies (tweets which directly respond to another tweet), geotagged (latitude/longitude coordinates associated with the tweet), user location (free text field which we were able to map to U.S. counties; see details below) and top languages (automatically detected language of the tweet). 
Retweets may or may not contain additional content created by the user doing the retweeting. 

Tweets also contain a large number of other metadata, such as user profile data and place information. 
User profiles contain information such as user handles, free text descriptions (often called ``bios"), and profile pictures. 
Places are named locations users decide to associate with a tweet. 
While Places describe physical locations, they do not necessarily imply that the tweet originated from this location. 
Twitter users may manually tag a location when their tweet is about that Place, regardless of the user's location at the time of posting. 
Due to the large number of additional fields available for each tweet, we do not provide counts for any additional content.

The monthly volume of each keyword is plotted in Figure \ref{fig:time series}. 
Here we plot the seven day running average of the total count (logged) of all tweets containing one of our keywords. 
We also label high profile events (e.g., deaths, court related events, and viral videos) which resulted in an increase in BLM related activity.
All labels marked with a single name indicate the date of police brutality-related killings.

In Figure \ref{fig:county plots} we visualized the spread of tweets across the United States over three equal time intervals: 2013 to 2015, 2016 to 2018, and 2019 to 2021.  Tweets are mapped to U.S. counties using tweet level latitude and longitude coordinates and self-report location information via a free text field in the user's profile. First, if a tweet object contains latitude and longitude coordinates, then the tweet can be trivially mapped to a U.S. county. Next, we examine the location free text field in the user profile and use a rule-based system to match this text to a list of unambiguous U.S. cities (i.e., New York City as opposed to Springfield) which can then be mapped to U.S. counties.  This process is described in \citet{schwartz2013characterizing}. 

Our data set consists of monthly csv files which contains a single row for each tweet. Rows consist of the numeric tweet id (status\_id; as given by the Twitter API) and three binary indicators for whether or not the tweet contains a \emph{BlackLivesMatter}, \emph{AllLivesMatter}, or \emph{BlueLivesMatter} related keyword (for four columns total). 

\subsection{LDA Topics}
The topic sets for each keyword contain three values: \emph{topic}, \emph{term}, and \emph{weight}. 
The \emph{topic} column is a numeric indicator for each topic: 1 through 100 for \emph{BlackLivesMatter}, 1 through 50 for \emph{AllLivesMatter}, and 1 through 25 for \emph{BlueLivesMatter}. 
The \emph{term} column is the word within the topic.
The \emph{weight} column is the conditional probability of the topic given the term, as derived through the LDA process. 
For each LDA topic set, we visualize the most prevalent topics across each corpus. 
To do this, we extract the relative frequency of single words (i.e., tokens) from each tweet in the no retweet, no reply, single keyword data sets described above: 10,881,298 \emph{BlackLivesMatter} tweets, 976,244 \emph{AllLivesMatter} tweets, and 1,069,362 \emph{BlueLivesMatter} tweets.
For each tweet we calculate the conditional probability of the topic given the tweet:

\begin{equation}
    P(\text{topic}|\text{tweet})=\sum_{\forall\text{token}\in\text{topic}}P(\text{topic}|\text{token})\times P(\text{token}|\text{tweet}).
\end{equation}

\begin{figure*}[!htbp]
\resizebox{1.0\textwidth}{!}{
\begin{tabular}{ccccc}
  \includegraphics[height=53cm]{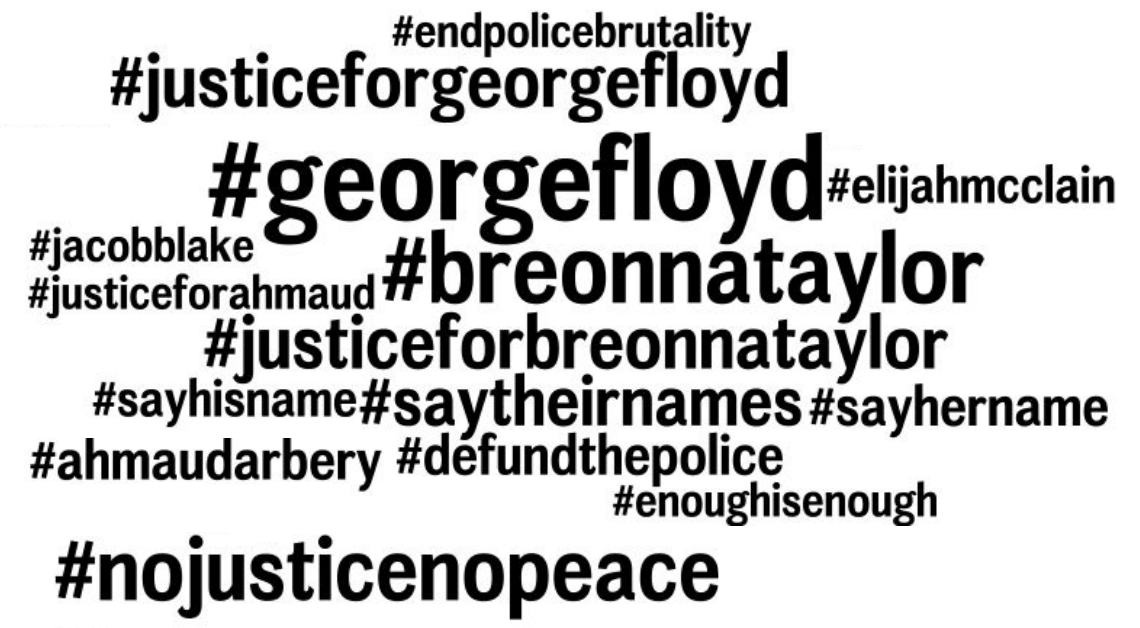} & 
  \includegraphics[height=53cm]{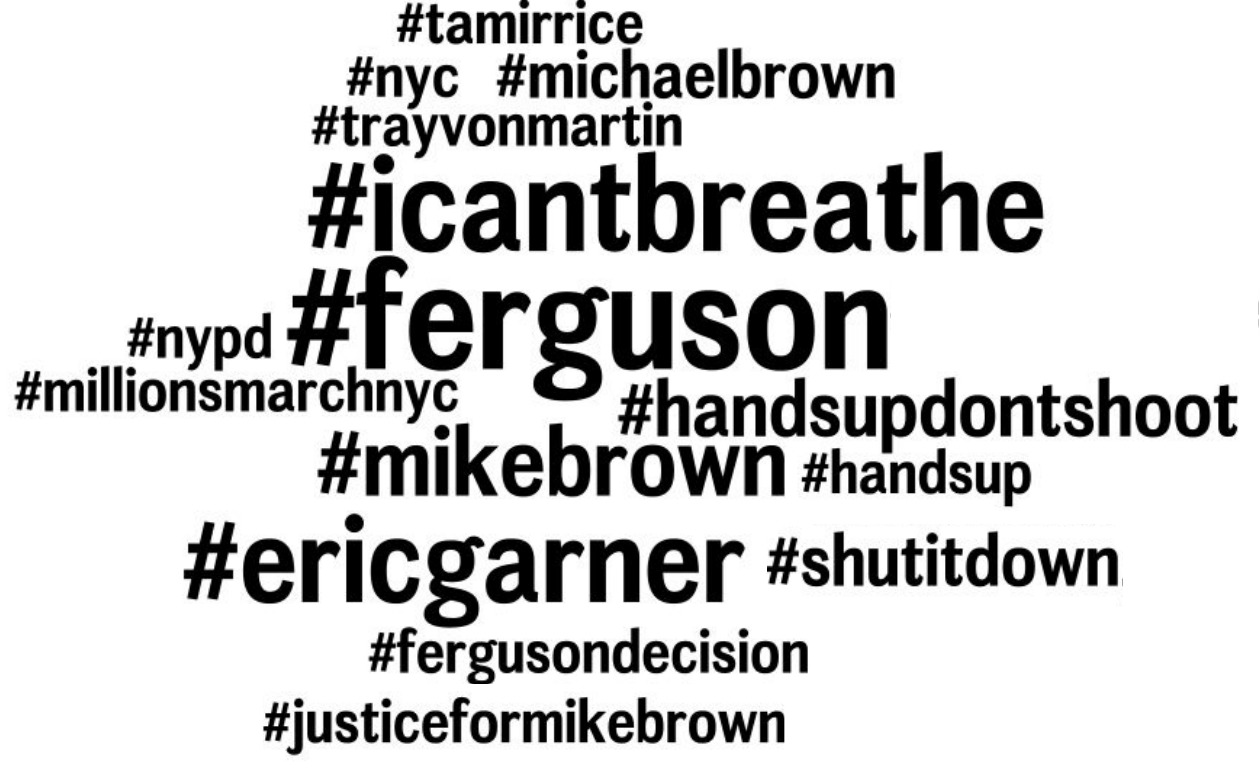} &
  \includegraphics[height=53cm]{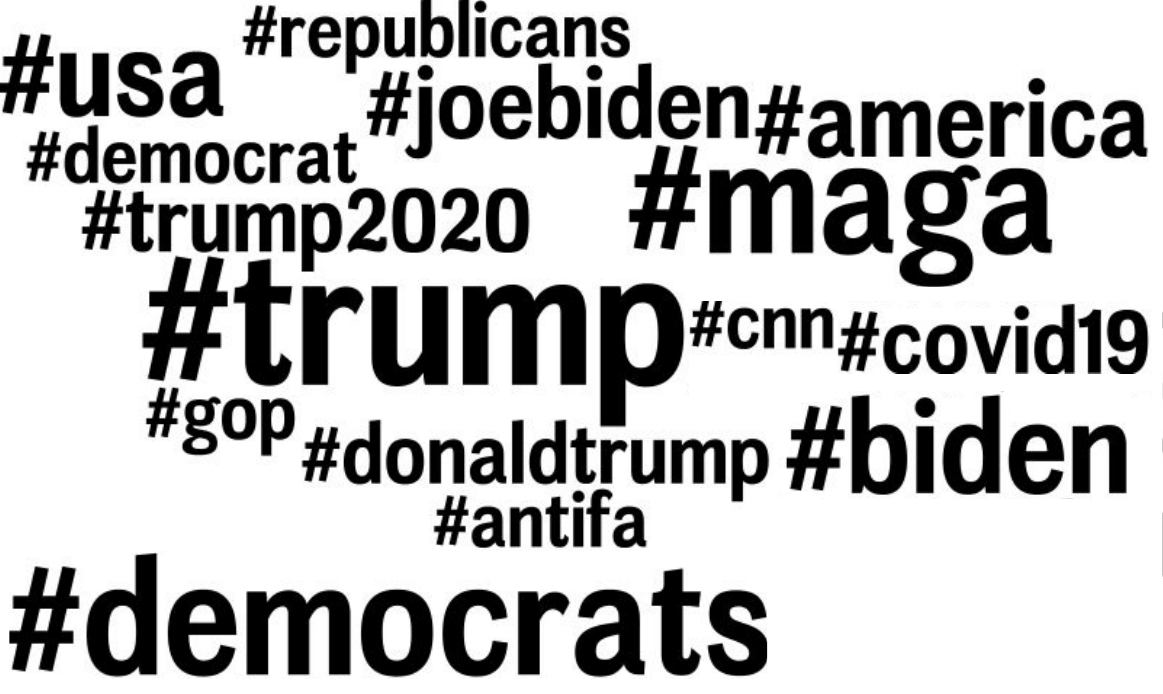} &
  \includegraphics[height=53cm]{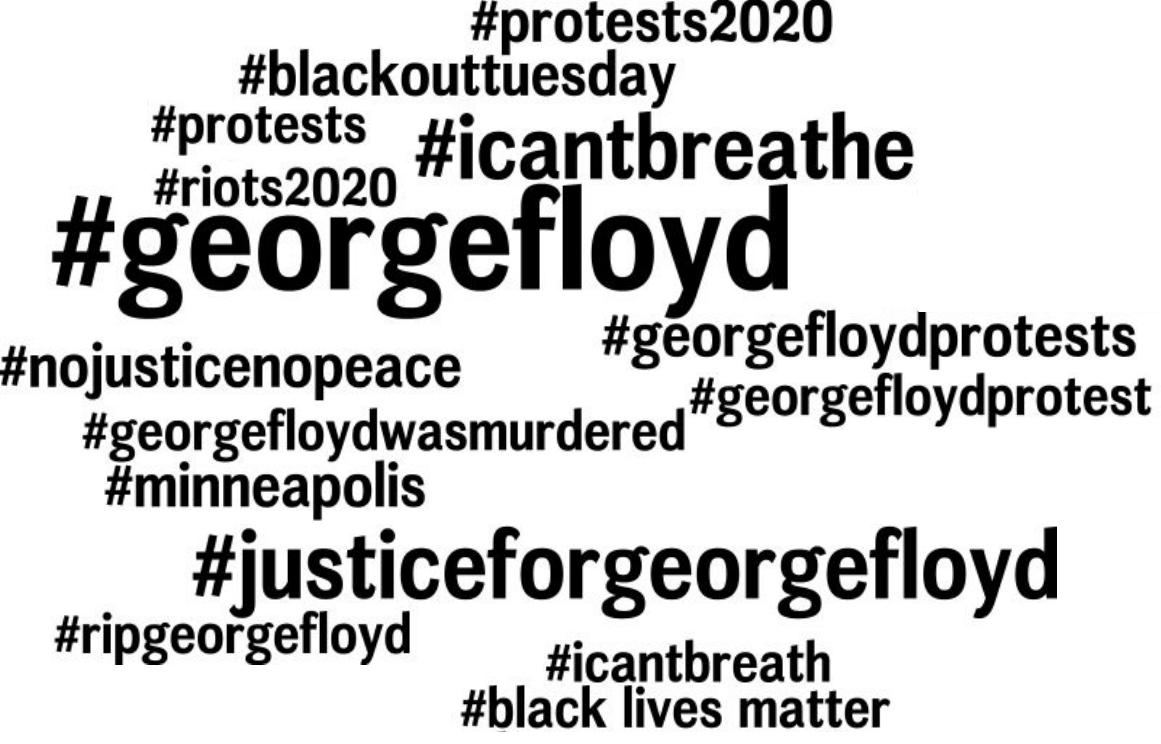} & \includegraphics[height=53cm]{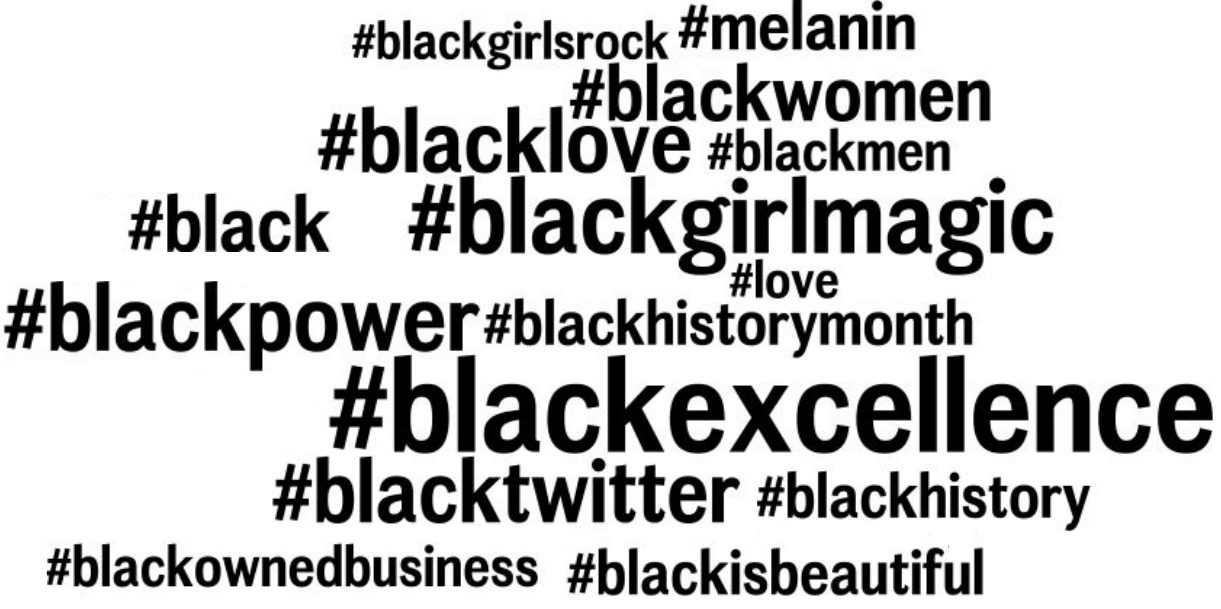} \\ 
  \\ \\ \newline

  \includegraphics[height=49cm]{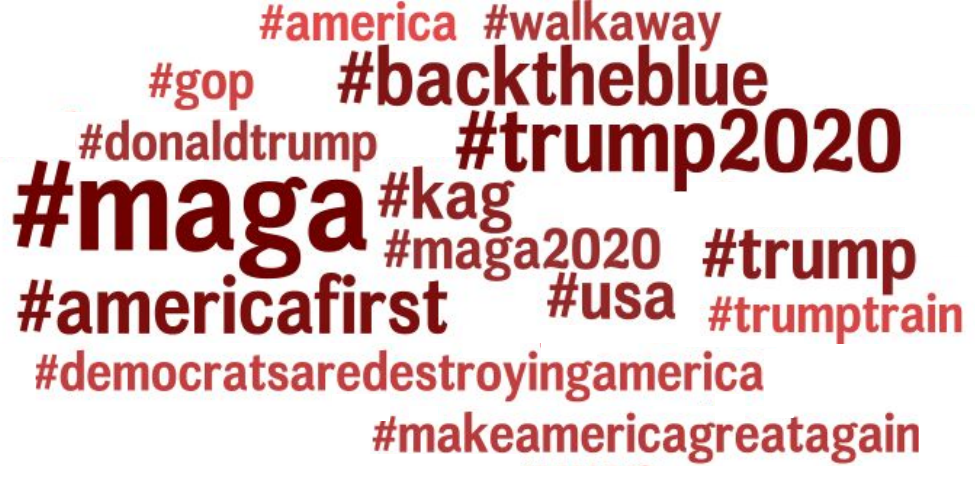} & 
  \includegraphics[height=49cm]{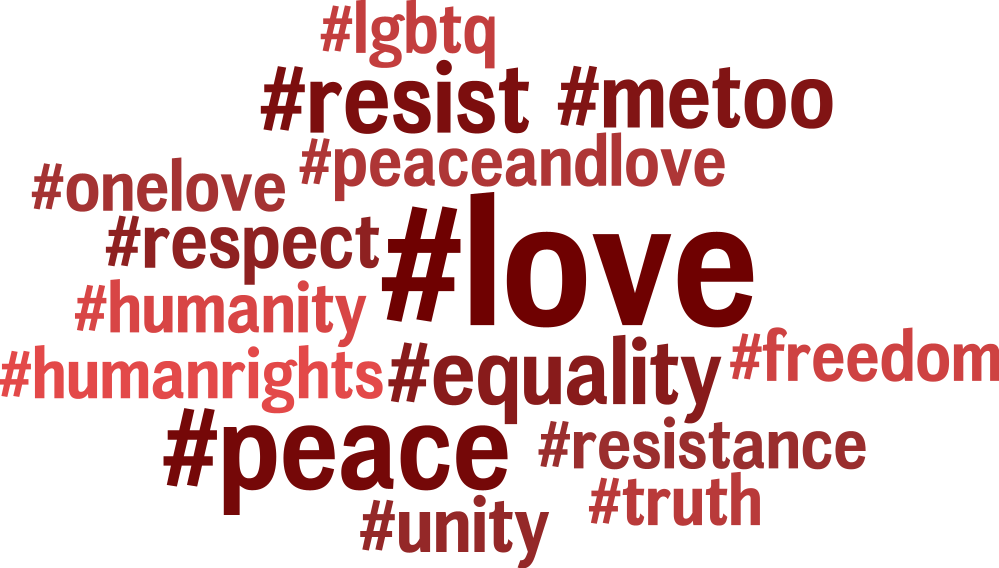} &
  \includegraphics[height=49cm]{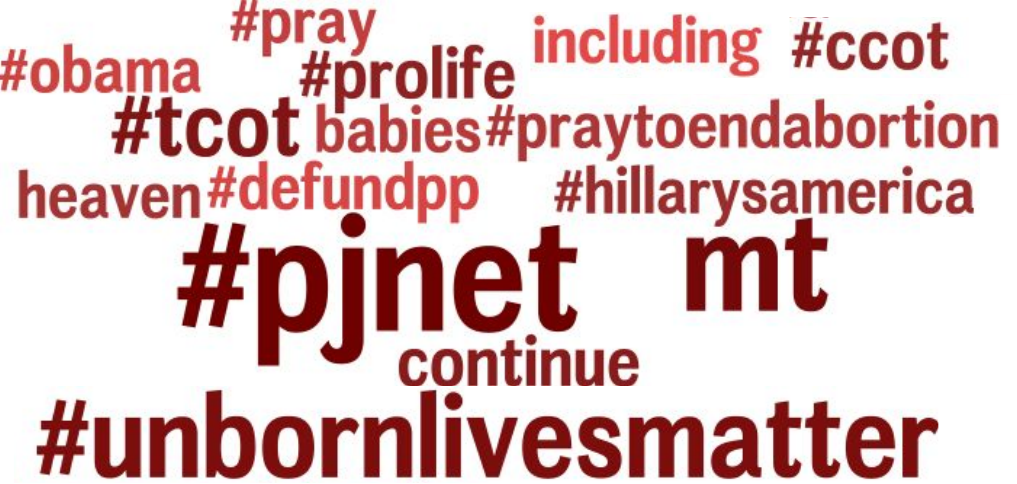} &
  \includegraphics[height=49cm]{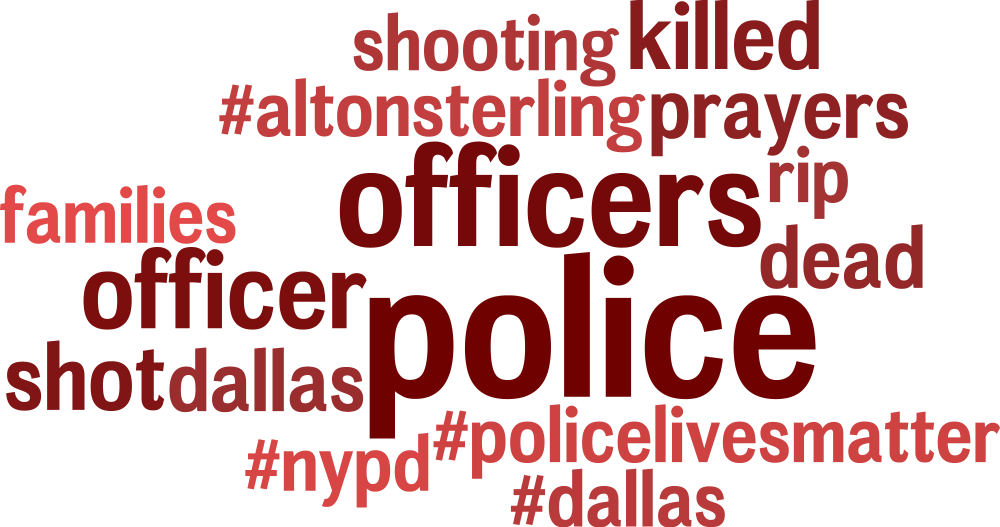} & \includegraphics[height=49cm]{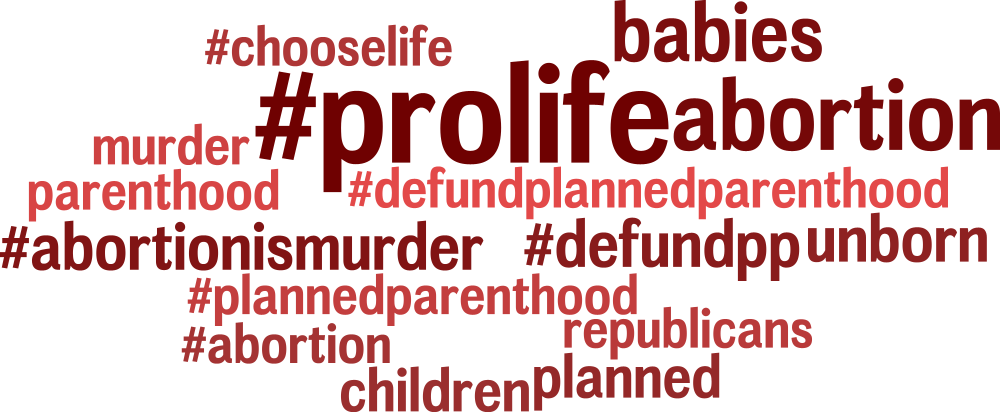} \\ 
  \\ \newline
  
  \includegraphics[height=53cm]{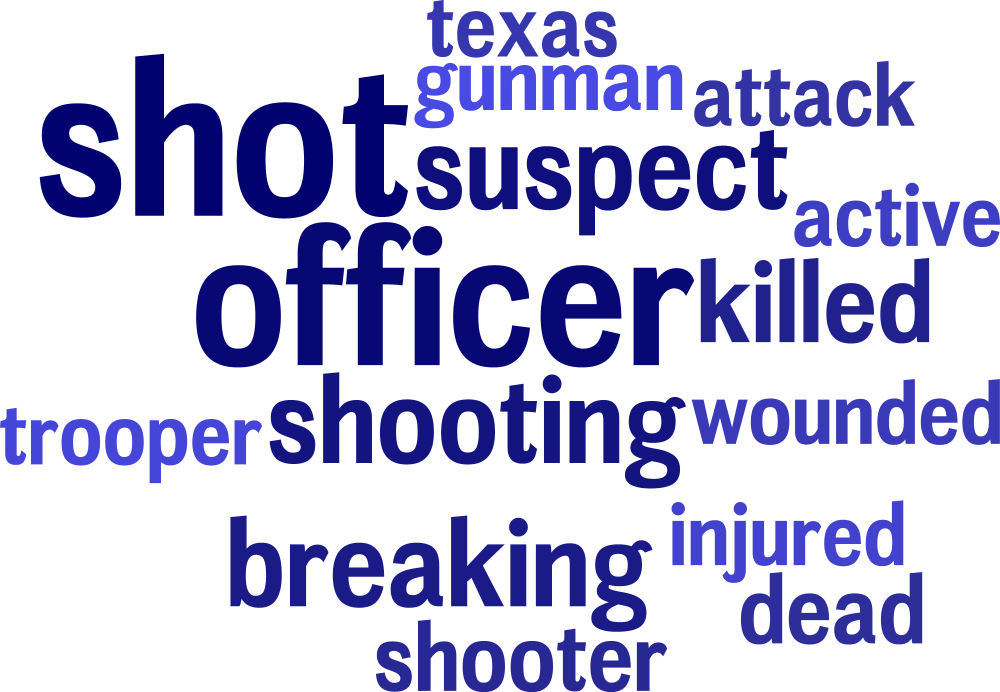} & 
  \includegraphics[height=53cm]{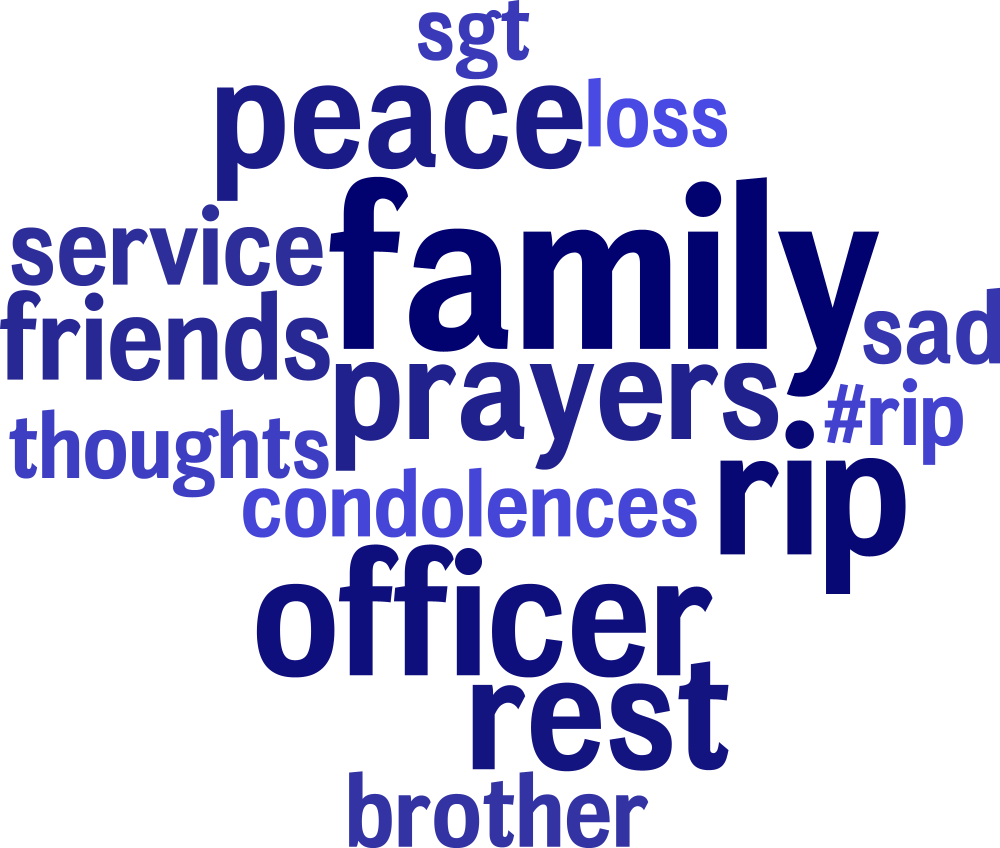} &
  \includegraphics[height=53cm]{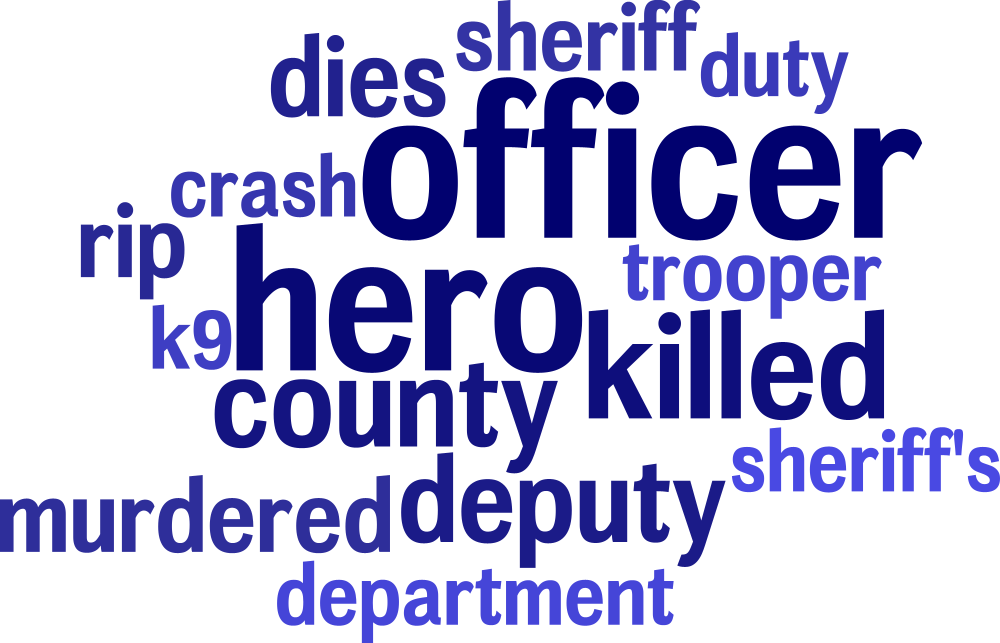} &
  \includegraphics[height=53cm]{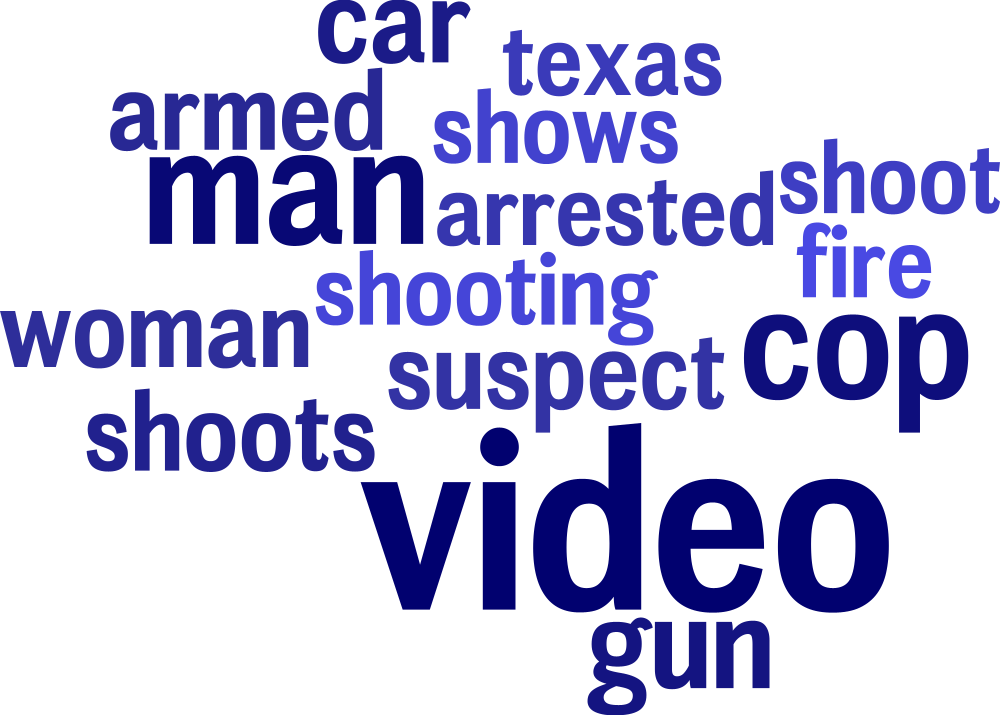} & \includegraphics[height=53cm]{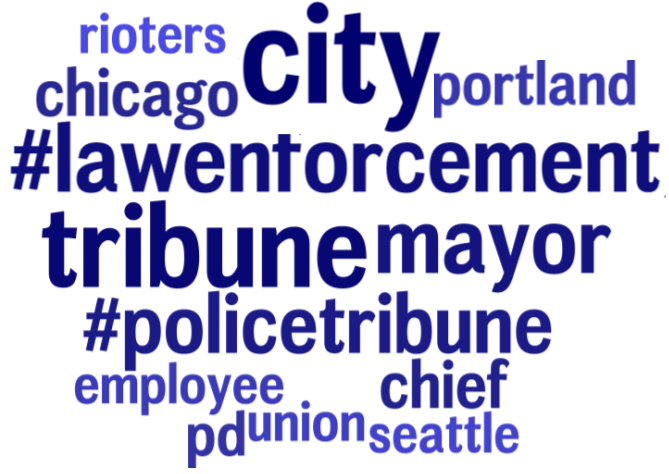} \\
\end{tabular}}
\caption{Word cloud visualizations of the top five most prevalent topics in the \textit{BlackLivesMatter} (colored in black), \textit{AllLivesMatter} (colored in red), and \textit{BlueLivesMatter} (colored in blue) data sets, respectively. Topics are ordered left to right by descending prevalence (average conditional probability of topic given tweet). Word clouds contain the 15 most prevalent words within the topic and words are sized according to their prevalence relative to other words in the topic.}
\label{fig:lda topics}
\end{figure*}

Here $P(\text{topic}|\text{token})$ is the conditional probability of the topic given the topic, which is estimated through the LDA process. 
We estimate $P(\text{token}|\text{tweet})$ as the relative frequency of the token given the tweet.  
Then for each tweet we have a conditional probability of each topic, which we average for each topic. 
Figure \ref{fig:lda topics} shows the 5 topics for each keyword with the highest average condition probability.

\section{Data Set Creation}
\label{sec:data set creation}

\subsection{Data Collection}
On July 14, 2016, we set up a data puller using the Python package TwitterMySQL\footnote{\label{twittermysql}\url{https://github.com/dlatk/TwitterMySQL}} to collect tweets matching at least one of our keywords: \emph{BlackLivesMatter}, \emph{AllLivesMatter} and \emph{BlueLivesMatter}. This package uses the official Twitter Application Programming Interface (API) to stream tweets in real time. The data puller continuously collected tweets from the Twitter stream until December 31, 2021. In total we collected 67,336,447 tweets. While the Twitter API was queried using the keywords \emph{BlackLivesMatter}, \emph{AllLivesMatter} and \emph{BlueLivesMatter}, the API delivers a more robust set of matching tweets. For example, a tweet might contain the phrase ``black lives matter”, ``blm" or ``\#blacklivesmatter", among other variations, instead of the exact keyword \emph{BlackLivesMatter}. 

We note that the Twitter API limits such streams to 1\% of the total Twitter volume at any given moment. To see if our keyword data set was limited at any point, we compared the monthly keyword volume to a full 1\% monthly pull (not limited to any single keyword, location, etc.) Our keyword data set pulled in a monthly average of 1,463,835 tweets (4,630,450 SD) as compared to a monthly average of 96,385,502 tweets (27.146,801 SD) from the 1\% pull. Since our data set is much smaller than the 1\% sample we do not believe our data set was limited by the Twitter API. 

Due to server maintenance, there were periods when we were unable to collect data. These include: October 17 through November 23, 2016; January 1 through January 21, 2017; March 11 through March 16, 2017; May 2 through December 18, 2018; and March 16 through March 20, 2019; June 1 through June 3, 2021; and November 19 through November 21, 2021. Additionally, the Black Lives Matter movement began in 2013, roughly three years before the beginning of our data collection. In order to fill these gaps, we used the Python package GetOldTweets3\footnote{\label{getoldtweets}\url{https://github.com/Mottl/GetOldTweets3}}, which pulls historical tweets containing a given keyword. Using this method, we collected 4,276,423 historical tweets across the dates listed above (i.e., the gaps in our data).

While Twitter data is publicly available, at any point a user may delete a tweet, delete their account, or set their account to private. Thus, when pulling prospective data, we collected tweets which may have been deleted or made private at some point after the initial pull. On the other hand, deleted or private tweets cannot be pulled with a retrospective collection. Thus, the number of tweets pulled prospectively or retrospectively can be very different, especially as one goes further back in time. In order to ensure the data set only contained presently available tweets, we executed a one-time historical. As a result, any tweet deleted after our initial pull will not be made available. Our final data set consisted of 63,884,799 tweets.

\begin{figure*}[ht!]
     \centering
     \begin{subfigure}[b]{0.32\textwidth}
         \centering
         \includegraphics[width=\textwidth]{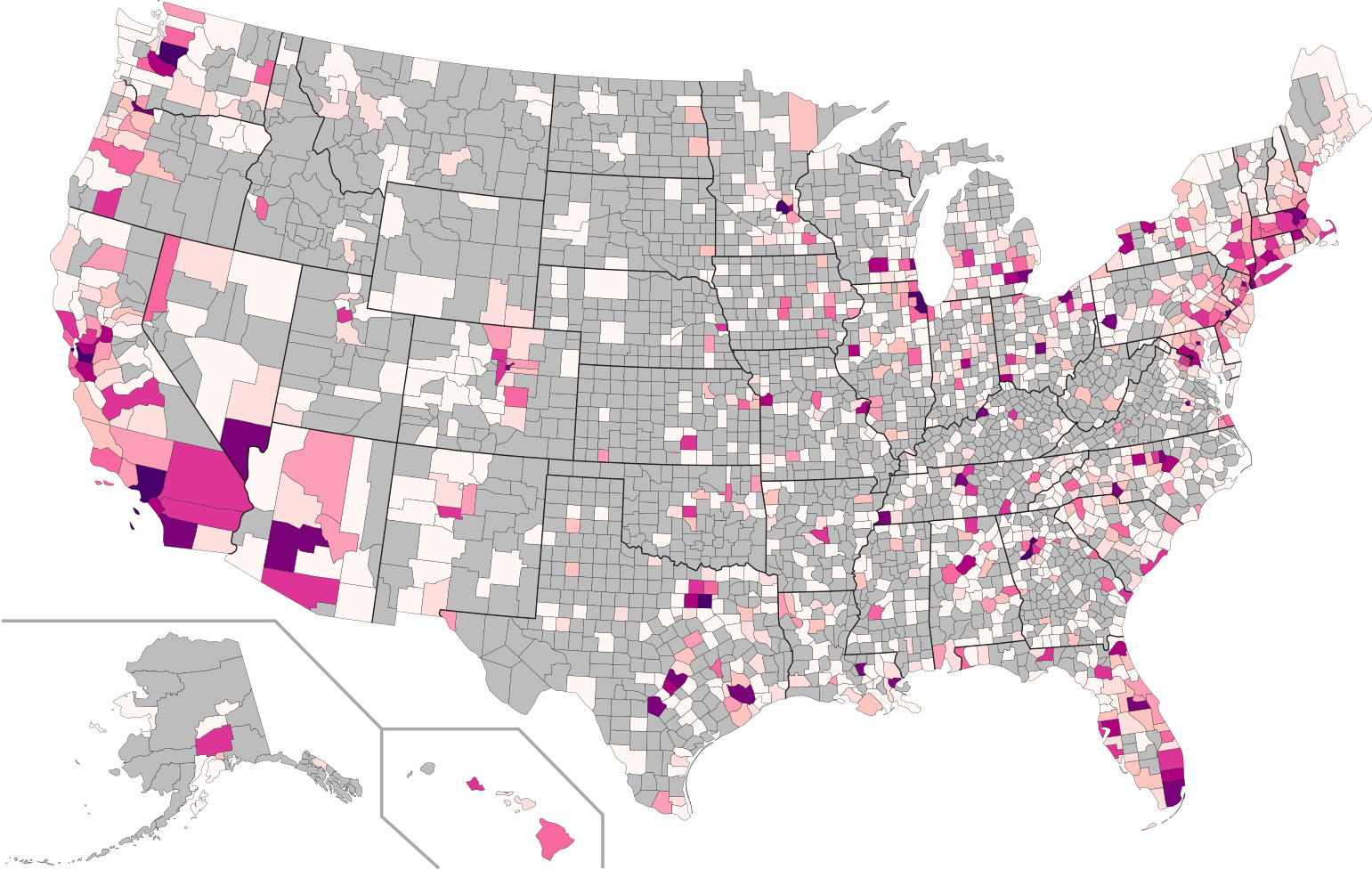}
         \caption{2013 - 2015}
         \label{fig:y equals x}
     \end{subfigure}
     \hfill
     \begin{subfigure}[b]{0.32\textwidth}
         \centering
         \includegraphics[width=\textwidth]{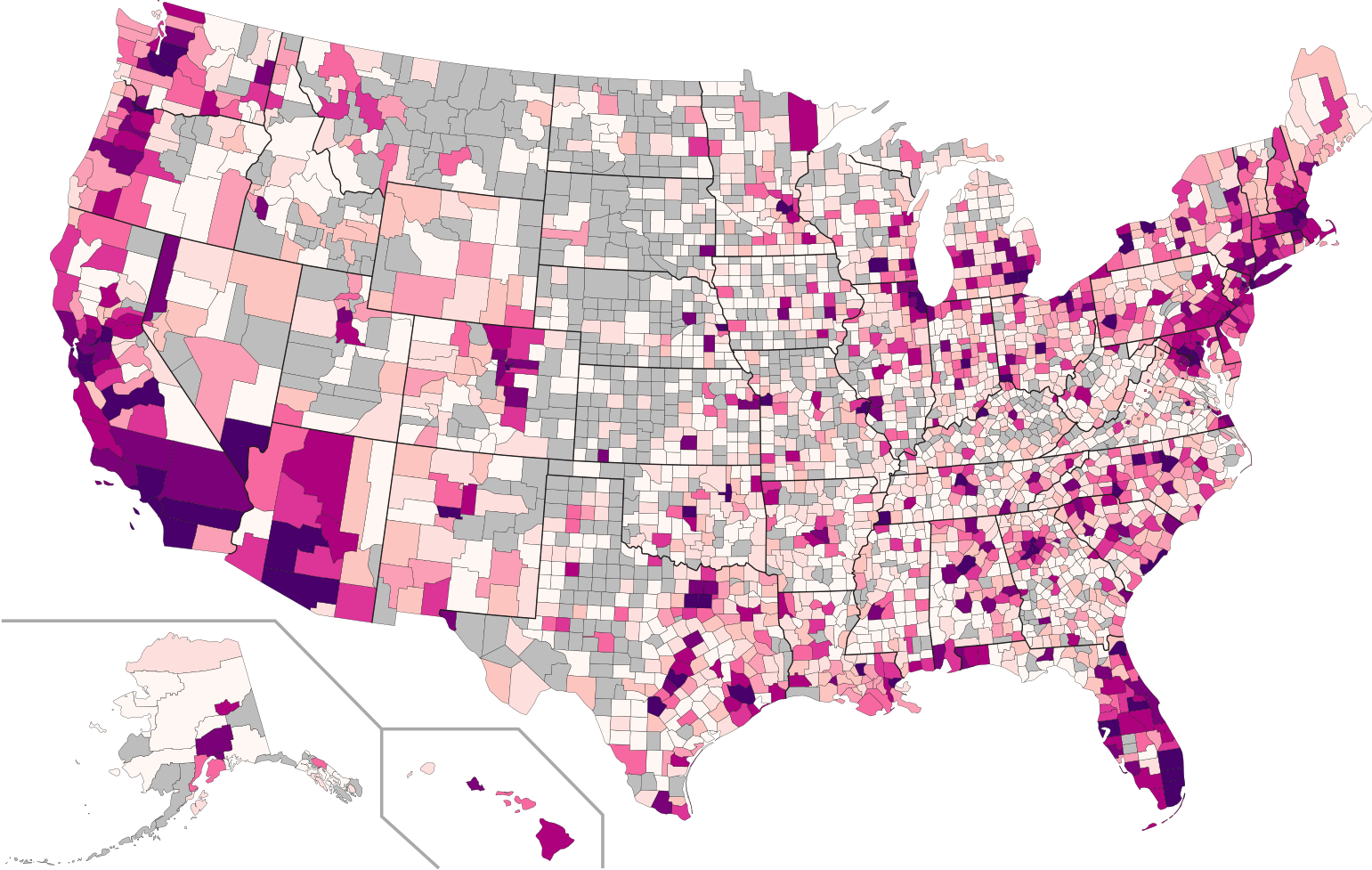}
         \caption{2016 - 2018}
         \label{fig:three sin x}
     \end{subfigure}
     \hfill
     \begin{subfigure}[b]{0.32\textwidth}
         \centering
         \includegraphics[width=\textwidth]{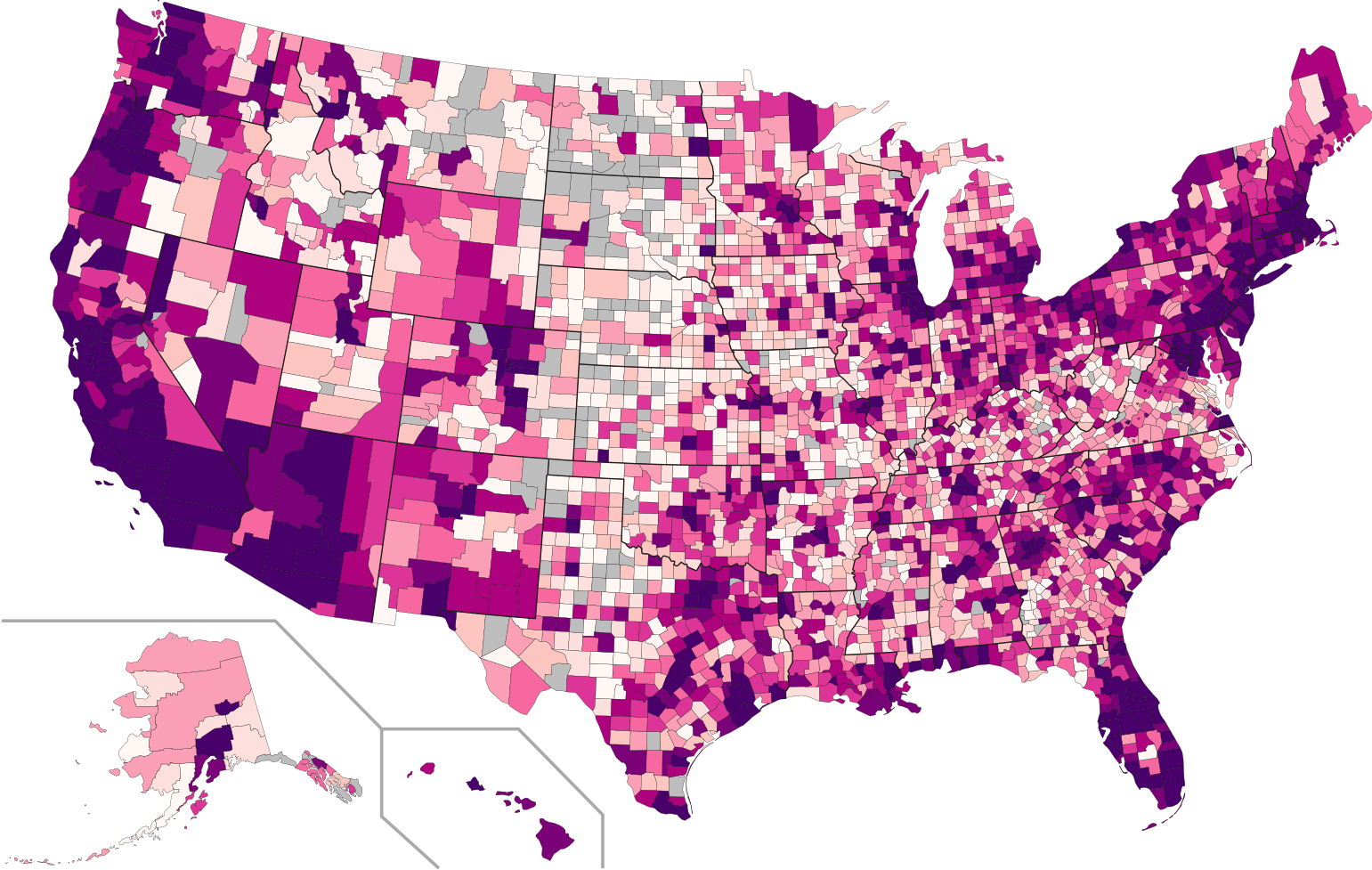}
         \caption{2019 - 2021}
         \label{fig:five over x}
     \end{subfigure}
     
     \begin{subfigure}[b]{0.4\textwidth}
         \centering
         \includegraphics[width=\textwidth]{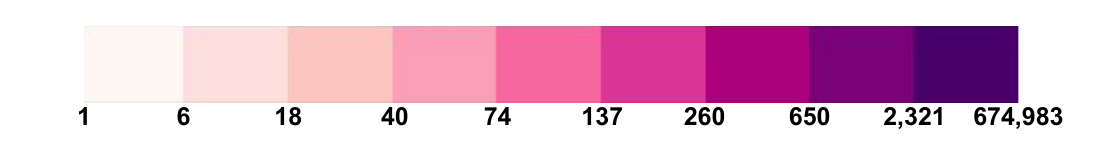}
     \end{subfigure}
        \caption{Distribution of \emph{BlackLivesMatter} tweets across the United States for three 3-year periods: 2013 to 2015, 2016 to 2018, and 2019 to 2021. Tweets are mapped to U.S. counties from latitude/longitude coordinates or self-reported user location. Counties grouped into 9 quantiles: darker shades indicate higher tweet density, lighter shades indicate lower tweet density, and grey areas contain no tweet data.}
        \label{fig:county plots}
\end{figure*}

\subsection{Topic Modeling}

For each keyword we created a set of topics using Latent Dirichlet Allocation (LDA; \citealp{blei2003latent}). 
LDA is a Bayesian mixture model which automatically groups together words that frequently appear in similar contexts. 
Topic models such as LDA are often used to statistically derive categories in a data driven fashion, rather than manually assigning words to predetermined categories. 
Given the highly specific nature of our data set, we created Content Specific LDA (CSLDA) topics~\cite{mz-2020-understanding}.
CSLDA is a method used for generating topics across a thematically narrow corpus (i.e., tweets about BLM as opposed to a random selection of tweets) and has successfully been used to model excessive drinking~\cite{giorgi2020cultural}, diabetes~\cite{griffis2020using}, and COVID-19~\cite{mz-2020-understanding} discussion on Twitter. 
In particular, CSLDA uses a text pre-processing step, executed before topic modeling, which identifies words most associated with the theme (i.e., Black Lives Matter). 
CSLDA does not assume that words frequently appearing in the keyword tweets are associated with the keyword.
For example, the retweet keyword ``RT" appears in a large number of our tweets, but is more associated with the Twitter platform than \emph{BlackLivesMatter}.
The CSLDA pipeline is briefly described below.
Further details on CSLDA can be found in Zamani et al.~\cite{mz-2020-understanding}.

In order to find words that are most associated with each keyword, we first built a corpus comprised of a random sample of tweets containing our keywords and a matched sample of tweets that do not. 
For each keyword, we randomly select 500,000 tweets that are neither replies nor retweets. 
We also select tweets containing only a single keyword, that is, no tweet contains some combination of \emph{BlackLivesMatter}, \emph{AllLivesMatter}, and \emph{BlueLivesMatter} keywords. 
For the matched sample, we randomly selected 500,000 tweets that do not contain any of the three keywords. 
These tweets were selected from a random 1\% stream of publicly available data and are also (1) not replies nor retweets, (2) written in English (as reported by Twitter's API), and (3) match the same temporal distribution as the random 500,000 \emph{BlackLivesMatter} tweets above (i.e., the number of tweets per year for the non-keyword set matches the number of tweets per year in the 500,000 \emph{BlackLivesMatter} tweets).
We then created three sets of one million tweets separately combining the random 500,000 random tweets with our three sets of 500,000 keyword tweets. 

Next, we broke up each tweet into its constituent words, in a process called tokenization.
As opposed to splitting up the tweets by white space, we use a tokenizer built specifically for social media text that can pick emojis, hashtags, and misspellings~\cite{schwartz2017dlatk}.
For each tweet, we then created a feature vector which records relative frequency of the words appearing in the tweet.
We also created a binary outcome for each tweet in the three matched data sets above: 1 if the tweet contains one of the keywords and 0 otherwise. 
Using this binary outcome and the feature vector, we calculated a weighted log odds ratio using an Informative Dirichlet prior~\cite{jurafsky2014narrative}.
This calculation estimates the difference in word frequency across the keyword data sets and their matched samples, using a prior which shrinks each keyword word frequency towards known frequencies in the matched sample.
In the end, we took the top 3,000, 2,500, and 2,500 associated unigrams (i.e., largest weighted log odds rations) for the \emph{BlackLivesMatter}, \emph{AllLivesMatter}, and \emph{BlueLivesMatter} keywords, respectively.
A larger number of unigrams associated with \emph{BlackLivesMatter} were chosen because the data set contains significantly more \emph{BlackLivesMatter} tweets than the \emph{AllLivesMatter} and \emph{BlueLivesMatter} data sets. 
Finally, we considered all tweets that are neither retweets nor replies and only contain a single keyword (i.e., no combination of \emph{BlackLivesMatter}, \emph{AllLivesMatter}, and \emph{BlueLivesMatter} keywords). 
These data sets contain 9,758,272; 1,386,087; and 1,167,006 tweets for \emph{BlackLivesMatter}, \emph{AllLivesMatter}, and \emph{BlueLivesMatter}, respectively.
We then filtered these tweets to contain only the unigrams derived in the previous step, removing words that do not differ in log odds frequency vs. the matched set.
We also removed urls, @-mentions, and all variations of the keywords (e.g., black, lives, matter, blm, \#blm, blacklivesmatter, and \#blacklivesmatter)
The LDA process was then run over these filtered data sets.

\begin{figure*}[ht!]
\centering
\minipage{0.9\textwidth}
  \includegraphics[width=\linewidth]{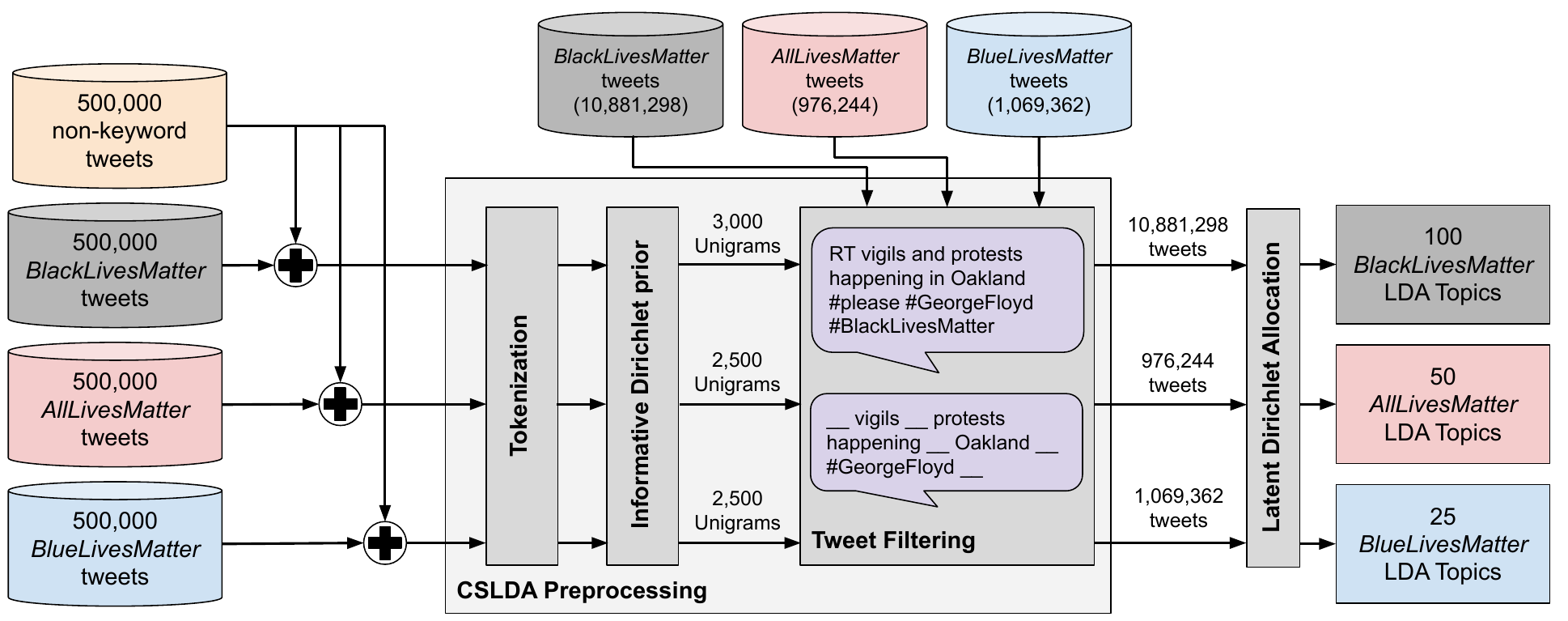}
  \caption{Content Specific LDA pipeline used for creating the three LDA topic sets. All tweets contained a single keyword (e.g., \emph{AllLivesMatter} keywords and not \emph{BlackLivesMatter} and \emph{BlueLivesMatter} keywords) and no retweets or replies were used.}
  \label{fig:cslda pipeline}
\endminipage
\end{figure*}

We used the Mallet software package,\footnote{\url{http://mallet.cs.umass.edu}} which uses Gibbs sampling~\cite{gelfand1990sampling} to estimate the latent variables of the topic. 
All default Mallet parameters are used except for $\alpha$, the prior on the expected number of topics per document. 
Here $\alpha=2$, since our tweets are shorter than typical documents (such as newspaper articles or blog posts), and thus contain fewer topics. 
This value has previously been used in CSLDA over Twitter data~\cite{giorgi2020cultural}.
For each keyword, we created three sets of LDA topics, varying the number of topics per set.
For \emph{BlackLivesMatter} we created sets of 25, 50, and 100 topics.
For both \emph{AllLivesMatter} and \emph{BlueLivesMatter} we created sets of 10, 25, and 50 topics, again noting that the total number of \emph{BlackLivesMatter} tweets are more than 10 times the number of tweets in the other two keywords data sets.
Other studies which have created LDA topics over thematically similar tweet data sets (e.g., COVID-19, excessive alcohol consumption, and maternal health) have used similarly sized topic sets~\cite{mz-2020-understanding,giorgi2020cultural,abebe2020quantifying}. Figure \ref{fig:cslda pipeline} shows the full pipeline.

In order to pick the final topic set, three of the authors qualitatively ranked the three sets. 
All raters were asked to consider (1) breadth of themes, (2) single topics contain single themes, and (3) themes are not repeated across a large number of topics.
A similar qualitative process was used to identify Twitter topics associated with maternal morality~\cite{abebe2020quantifying}.
All three raters agreed on the 100 topic set for \emph{BlackLivesMatter} and the 50 topic set for \emph{AllLivesMatter}. 
The three raters did not initially agree on the \emph{BlueLivesMatter} topic set (with raters split between the 25 and 50 topic sets), but eventually settled on the 25 topic set since themes repeated across a large number of topics in the 50 topic set.

To evaluate the quality of the topics, we use two metrics. The first metric, Topic Uniqueness (TU), is a measure of diversity~\cite{nan2019topic}. TU is inversely proportional to the number of times a set of $L$ representative keywords is repeated across the set of $K$ topics, with high TU meaning the words are rarely repeated, and thus the topics are unique. 
Traditionally, TU scores are between 1 and $1/K$, where $K$ is the number of topics. Since the number of topics across our three keywords varies, we normalize the TU score to be between 0 and 1. We set $L=30$ and $K$ equal to 100, 50, and 25, for the \emph{BlackLivesMatter}, \emph{AllLivesMatter}, and \emph{BlueLivesMatter} topic sets, respectively. This produces TU scores of .79 for \emph{BlackLivesMatter}, .97 for \emph{AllLivesMatter}, and 1.0 for \emph{BlueLivesMatter}. We note that as $L$ increases, TU scores should decreases since the probability of a given word appearing in more than one topic will increase. Traditionally, $L$ is set to 10, which we increase to 30 in order to give a more conservative estimate~\cite{nan2019topic}. 

The second metric measures coherence, or the semantic similarity between the words in the topic, using Normalized Pointwise Mutual Information (NPMI)~\cite{syed2017full}. Coherence is calculated for each topic and then averaged across all topics within each keyword topic set. Scores range between 0 and 1, where topics with high semantic similarity between words having scores closer to 1. We use the Gensim Python package to calculate these scores~\cite{rehurek_lrec} and see scores of 0.64 for \emph{BlackLivesMatter}, 0.73 for \emph{AllLivesMatter}, and 0.70 for \emph{BlueLivesMatter}.

\begin{table}[b]
\begin{tabular}{lccc} \toprule
 & \begin{tabular}[c]{@{}c@{}}Number\\ of Topics\end{tabular} & \begin{tabular}[c]{@{}c@{}}Topic \\ Uniqueness\end{tabular} & Coherence \\
 \hline
BlackLivesMatter & 100 & 1.00 & 0.64 \\
AllLivesMatter & 50 & 0.97 & 0.73 \\
BlueLivesMatter & 25 & 0.79 & 0.70 \\ \bottomrule 
\end{tabular}
\caption{Topic quality as measured by Topic Uniqueness and coherence.}
\label{tab:my-table}
\end{table}

\section{Usage Notes}

Due to Twitter’s Terms of Service, only numeric tweet IDs can be publicly shared. The numeric IDs can be used to pull the full tweet set using the Twitter API. There are a number of open source software packages which allow researchers to easily interface with the API. 
The authors used the Python package TwitterMySQL$^1$, which saves tweet information in a MySQL database. 
Other packages exist which do not rely on relational databases, such as the Python package twarc\footnote{\label{twarc}\url{https://github.com/DocNow/twarc}}, which saves tweets to text files in JSON format, or Hydrator\footnote{\label{hydrator}\url{https://github.com/DocNow/hydrator}}, which relies on an easy to use GUI and saves tweets to both JSON and CSV formats. 
Regardless of which tool is used to download the Twitter data, researchers need an active Twitter Developer account in order to access their API.

\section{Code Availability}

The data sets have been created using Python 3.5 and MySQL 5.5. The code is available through GitHub\footnote{\url{https://github.com/sjgiorgi/blm_twitter_corpus}}. 
A Python script and short tutorial have been developed to aid in obtaining this data from the Twitter API. 

\section{Conclusions}
To date, this data set is the largest publicly available collection of Black Lives Matter related social media posts. This data set was created to aid researchers in studying social media activity and discourse around the Black Lives Matter movement, with no specific task in mind. It is our hope that a central repository for this large, multi-year data base will give researchers easier access to the data, especially those researchers less comfortable using open source APIs or who lack computational bandwidth or storage capacity. 

We believe there are a number of applications and potential use cases for this data, which could include analyses of conversations, temporal shifts, and spatial trends. From this data set, one could pull conversations associated with each tweets (i.e., replies and retweets) via the Twitter API. Using this larger data set one could examine conversations within and between BLM and the counter protests. Additionally, one could compare conversations before and after the murder of George Floyd, since there was a major increase in Twitter activity after that event.

Using location data associated with the tweets and Twitter users, one could examine how BLM social media activity and conversations have evolved spatially. Figure \ref{fig:county plots} shows how the movement has grown within the U.S. and we note that the data set contains tweets from over 100 countries. One could attempt to look at spatial or cultural components of BLM conversations.

Additionally, one could combine the Twitter data with other geolocated data such as demographics, socioeconomics, and voting behavior. Past studies have examined BLM related protests and police-caused deaths~\cite{williamson2018black}, both of which could be combined with this Twitter data. Within the U.S., there are a number of publicly available data sets which measure racial attitudes at the population level. One such data set is from Project Implicit\footnote{\url{https://osf.io/y9hiq/}} which has publicly released data from the race implicit association test (IAT; \citealp{greenwald1998measuring}). This and other similar data sets can be mapped to U.S. locations such as counties and then correlated with the BLM Twitter data. A similar analysis was carried out by \citet{sawyer2018implicit}, who examined implicit and explicit attitudes during various phases of the BLM movement.

\section{Ethics Statement}
Here we consider the ethical questions outlined in Datasheets for Datasets~\cite{gebru2021datasheets}. As with any Twitter data set, there are a number of ethical concerns, especially due to the nature of this data set (i.e., racial justice and grassroots social movements). First, it is possible to identify individuals within the data set, though we note that this is only possible after rehydrating the tweet set and only because the data is publicly available (i.e., public Twitter profiles with public tweets). Thus, an individual would have to have posted something identifying on their account at some point in time that remains public at the time of rehydration. Along with any Twitter data, there are other possibly sensitive attributes such as images, location information, and friend/follower networks. Due to the nature of the data set, this could also potentially identify a person's support or opposition to a political movement. Finally, no individuals within the data set have consented to the authors to have their data shared, though, again, we note that the data used in this article is publicly available and distributed within Twitter’s Terms of Services (i.e., only numeric tweet IDs are distributed). While there is no official way to ``revoke consent", Twitter users may delete tweets, delete accounts, or set accounts to private, at which point any tweet in our data set would no longer be available. It was our hope that repulling all data at the time of writing would ensure the numbers reported within reflect the most current publicly available version of the data set.  

While we have released tweets related to counter protests (All Lives Matter and Blue Lives Matter), the authors do not intend to draw equivalences between the counter protests and Black Lives Matter. Despite the fact that all three protests are given equal space in the analysis, we believe the numbers reported show a different story: an overwhelming majority of tweets are related to \emph{BlackLivesMatter} as opposed to the other protests. 

It is our hope that this data set is used for social good, though there are a number of questionable use cases such as monitoring or forecasting of protests by law enforcement. As such, we limit the distribution of this data to non-commercial, research entities in hopes of limiting surveillance type tasks by for-profit entities.


\section{Acknowledgments}
This research was supported in part by the Intramural Research Program of the NIH, National Institute on Drug Abuse (NIDA).

\bibliography{references.bib}

\begin{thebibliography}{40}
\providecommand{\natexlab}[1]{#1}

\bibitem[{Abebe et~al.(2020)Abebe, Giorgi, Tedijanto, Buffone, and
  Schwartz}]{abebe2020quantifying}
Abebe, R.; Giorgi, S.; Tedijanto, A.; Buffone, A.; and Schwartz, H. A.~A. 2020.
\newblock Quantifying Community Characteristics of Maternal Mortality Using
  Social Media.
\newblock In \emph{Proceedings of The Web Conference 2020}, 2976--2983.

\bibitem[{Aceves(2018)}]{aceves2018virtual}
Aceves, W.~J. 2018.
\newblock Virtual hatred: How Russia tried to start a race war in the united
  states.
\newblock \emph{Mich. J. Race \& L.}, 24: 177.

\bibitem[{Blei, Ng, and Jordan(2003)}]{blei2003latent}
Blei, D.~M.; Ng, A.~Y.; and Jordan, M.~I. 2003.
\newblock Latent dirichlet allocation.
\newblock \emph{Journal of machine Learning research}, 3(Jan): 993--1022.

\bibitem[{Blevins et~al.(2019)Blevins, Lee, McCabe, and
  Edgerton}]{blevins2019tweeting}
Blevins, J.~L.; Lee, J.~J.; McCabe, E.~E.; and Edgerton, E. 2019.
\newblock Tweeting for social justice in\# Ferguson: Affective discourse in
  Twitter hashtags.
\newblock \emph{new media \& society}, 21(7): 1636--1653.

\bibitem[{Bor et~al.(2018)Bor, Venkataramani, Williams, and
  Tsai}]{bor2018police}
Bor, J.; Venkataramani, A.~S.; Williams, D.~R.; and Tsai, A.~C. 2018.
\newblock Police killings and their spillover effects on the mental health of
  black Americans: a population-based, quasi-experimental study.
\newblock \emph{The Lancet}, 392(10144): 302--310.

\bibitem[{Collaborators et~al.(2021)}]{gbd2021fatal}
Collaborators, G. . P. V. U.~S.; et~al. 2021.
\newblock Fatal police violence by race and state in the USA, 1980--2019: a
  network meta-regression.
\newblock \emph{The Lancet}, 398(10307): 1239--1255.

\bibitem[{Edwards, Lee, and Esposito(2019)}]{edwards2019risk}
Edwards, F.; Lee, H.; and Esposito, M. 2019.
\newblock Risk of being killed by police use of force in the United States by
  age, race--ethnicity, and sex.
\newblock \emph{Proceedings of the National Academy of Sciences}, 116(34):
  16793--16798.

\bibitem[{Eichstaedt et~al.(2021)Eichstaedt, Sherman, Giorgi, Roberts,
  Reynolds, Ungar, and Guntuku}]{eichstaedt2021emotional}
Eichstaedt, J.~C.; Sherman, G.~T.; Giorgi, S.; Roberts, S.~O.; Reynolds, M.~E.;
  Ungar, L.~H.; and Guntuku, S.~C. 2021.
\newblock The emotional and mental health impact of the murder of George Floyd
  on the US population.
\newblock \emph{Proceedings of the National Academy of Sciences}, 118(39).

\bibitem[{Gallagher et~al.(2018)Gallagher, Reagan, Danforth, and
  Dodds}]{gallagher2018divergent}
Gallagher, R.~J.; Reagan, A.~J.; Danforth, C.~M.; and Dodds, P.~S. 2018.
\newblock Divergent discourse between protests and counter-protests:
  \#BlackLivesMatter and \#AllLivesMatter.
\newblock \emph{PloS one}, 13(4): e0195644.

\bibitem[{Galovski et~al.(2016)Galovski, Peterson, Beagley, Strasshofer, Held,
  and Fletcher}]{galovski2016exposure}
Galovski, T.~E.; Peterson, Z.~D.; Beagley, M.~C.; Strasshofer, D.~R.; Held, P.;
  and Fletcher, T.~D. 2016.
\newblock Exposure to violence during Ferguson protests: Mental health effects
  for law enforcement and community members.
\newblock \emph{Journal of Traumatic Stress}, 29(4): 283--292.

\bibitem[{Gebru et~al.(2021)Gebru, Morgenstern, Vecchione, Vaughan, Wallach,
  Iii, and Crawford}]{gebru2021datasheets}
Gebru, T.; Morgenstern, J.; Vecchione, B.; Vaughan, J.~W.; Wallach, H.; Iii,
  H.~D.; and Crawford, K. 2021.
\newblock Datasheets for datasets.
\newblock \emph{Communications of the ACM}, 64(12): 86--92.

\bibitem[{Gelfand and Smith(1990)}]{gelfand1990sampling}
Gelfand, A.~E.; and Smith, A.~F. 1990.
\newblock Sampling-based approaches to calculating marginal densities.
\newblock \emph{Journal of the American statistical association}, 85(410):
  398--409.

\bibitem[{Giorgi et~al.(2022)Giorgi, Guntuku, Himelein-Wachowiak, Kwarteng,
  Hwang, Rahman, and Curtis}]{salvatore_giorgi_2022_5835260}
Giorgi, S.; Guntuku, S.~C.; Himelein-Wachowiak, M.; Kwarteng, A.; Hwang, S.;
  Rahman, M.; and Curtis, B. 2022.
\newblock {Twitter Data of the \#BlackLivesMatter Movement And Counter
  Protests: 2013 to 2021}.

\bibitem[{Giorgi et~al.(2020)Giorgi, Yaden, Eichstaedt, Ashford, Buffone,
  Schwartz, Ungar, and Curtis}]{giorgi2020cultural}
Giorgi, S.; Yaden, D.~B.; Eichstaedt, J.~C.; Ashford, R.~D.; Buffone, A.~E.;
  Schwartz, H.~A.; Ungar, L.~H.; and Curtis, B. 2020.
\newblock Cultural Differences in Tweeting about Drinking Across the US.
\newblock \emph{International Journal of Environmental Research and Public
  Health}, 17(4): 1125.

\bibitem[{Giorgi et~al.(2021)Giorgi, Zavarella, Tanev, Stefanovitch, Hwang,
  Hettiarachchi, Ranasinghe, Kalyan, Tan, Tan, Andrews, Hu, Stoehr, Re, Vegh,
  Atzenhofer, Curtis, and
  H{\"u}rriyeto{\u{g}}lu}]{giorgi-etal-2021-discovering}
Giorgi, S.; Zavarella, V.; Tanev, H.; Stefanovitch, N.; Hwang, S.;
  Hettiarachchi, H.; Ranasinghe, T.; Kalyan, V.; Tan, P.; Tan, S.; Andrews, M.;
  Hu, T.; Stoehr, N.; Re, F.~I.; Vegh, D.; Atzenhofer, D.; Curtis, B.; and
  H{\"u}rriyeto{\u{g}}lu, A. 2021.
\newblock Discovering Black Lives Matter Events in the {U}nited {S}tates:
  Shared Task 3, {CASE} 2021.
\newblock In \emph{Proceedings of the 4th Workshop on Challenges and
  Applications of Automated Extraction of Socio-political Events from Text
  (CASE 2021)}, 218--227. Online: Association for Computational Linguistics.

\bibitem[{Greenwald, McGhee, and Schwartz(1998)}]{greenwald1998measuring}
Greenwald, A.~G.; McGhee, D.~E.; and Schwartz, J.~L. 1998.
\newblock Measuring individual differences in implicit cognition: the implicit
  association test.
\newblock \emph{Journal of personality and social psychology}, 74(6): 1464.

\bibitem[{Griffis et~al.(2020)Griffis, Asch, Schwartz, Ungar, Buttenheim, Barg,
  Mitra, and Merchant}]{griffis2020using}
Griffis, H.; Asch, D.~A.; Schwartz, H.~A.; Ungar, L.; Buttenheim, A.~M.; Barg,
  F.~K.; Mitra, N.; and Merchant, R.~M. 2020.
\newblock Using Social Media to Track Geographic Variability in Language About
  Diabetes: Infodemiology Analysis.
\newblock \emph{JMIR diabetes}, 5(1): e14431.

\bibitem[{Horowitz(2020)}]{horowitz2020amid}
Horowitz, J. 2020.
\newblock \emph{Amid protests, majorities across racial and ethnic groups
  express support for the Black Lives Matter movement}.
\newblock Pew Research Center.

\bibitem[{H{\"u}rriyeto{\u{g}}lu et~al.(2021)H{\"u}rriyeto{\u{g}}lu, Tanev,
  Zavarella, Piskorski, Yeniterzi, Mutlu, Yuret, and
  Villavicencio}]{hurriyetoglu-etal-2021-challenges}
H{\"u}rriyeto{\u{g}}lu, A.; Tanev, H.; Zavarella, V.; Piskorski, J.; Yeniterzi,
  R.; Mutlu, O.; Yuret, D.; and Villavicencio, A. 2021.
\newblock Challenges and Applications of Automated Extraction of
  Socio-political Events from Text ({CASE} 2021): Workshop and Shared Task
  Report.
\newblock In \emph{Proceedings of the 4th Workshop on Challenges and
  Applications of Automated Extraction of Socio-political Events from Text
  (CASE 2021)}, 1--9. Online: Association for Computational Linguistics.

\bibitem[{Ince, Rojas, and Davis(2017)}]{ince2017social}
Ince, J.; Rojas, F.; and Davis, C.~A. 2017.
\newblock The social media response to Black Lives Matter: how Twitter users
  interact with Black Lives Matter through hashtag use.
\newblock \emph{Ethnic and racial studies}, 40(11): 1814--1830.

\bibitem[{Jurafsky et~al.(2014)Jurafsky, Chahuneau, Routledge, and
  Smith}]{jurafsky2014narrative}
Jurafsky, D.; Chahuneau, V.; Routledge, B.~R.; and Smith, N.~A. 2014.
\newblock Narrative framing of consumer sentiment in online restaurant reviews.
\newblock \emph{First Monday}.

\bibitem[{Keib, Himelboim, and Han(2018)}]{keib2018important}
Keib, K.; Himelboim, I.; and Han, J.-Y. 2018.
\newblock Important tweets matter: Predicting retweets in the\#
  BlackLivesMatter talk on twitter.
\newblock \emph{Computers in Human Behavior}, 85: 106--115.

\bibitem[{Kishi and Jones(2020)}]{kishi2020demonstrations}
Kishi, R.; and Jones, S. 2020.
\newblock Demonstrations \& Political Violence in America: New Data for Summer
  2020.
\newblock \emph{The Armed Conflict Location \& Event Data Project (ACLED)}.

\bibitem[{Mundt, Ross, and Burnett(2018)}]{mundt2018scaling}
Mundt, M.; Ross, K.; and Burnett, C.~M. 2018.
\newblock Scaling social movements through social media: The case of black
  lives matter.
\newblock \emph{Social Media+ Society}, 4(4): 2056305118807911.

\bibitem[{Nan et~al.(2019)Nan, Ding, Nallapati, and Xiang}]{nan2019topic}
Nan, F.; Ding, R.; Nallapati, R.; and Xiang, B. 2019.
\newblock Topic Modeling with Wasserstein Autoencoders.
\newblock In \emph{Proceedings of the 57th Annual Meeting of the Association
  for Computational Linguistics}, 6345--6381.

\bibitem[{Putnam, Chenoweth, and Pressman(2020)}]{putnam2020floyd}
Putnam, L.; Chenoweth, E.; and Pressman, J. 2020.
\newblock The Floyd Protests are the broadest in US history—and are spreading
  to white, small-town America.
\newblock \emph{Washington Post}, 6.

\bibitem[{{\v R}eh{\r u}{\v r}ek and Sojka(2010)}]{rehurek_lrec}
{\v R}eh{\r u}{\v r}ek, R.; and Sojka, P. 2010.
\newblock {Software Framework for Topic Modelling with Large Corpora}.
\newblock In \emph{{Proceedings of the LREC 2010 Workshop on New Challenges for
  NLP Frameworks}}, 45--50. Valletta, Malta: ELRA.

\bibitem[{Ross(2015)}]{ross2015multi}
Ross, C.~T. 2015.
\newblock A multi-level Bayesian analysis of racial bias in police shootings at
  the county-level in the United States, 2011--2014.
\newblock \emph{PloS one}, 10(11): e0141854.

\bibitem[{Sawyer and Gampa(2018)}]{sawyer2018implicit}
Sawyer, J.; and Gampa, A. 2018.
\newblock Implicit and explicit racial attitudes changed during Black Lives
  Matter.
\newblock \emph{Personality and Social Psychology Bulletin}, 44(7): 1039--1059.

\bibitem[{Schwartz(2020)}]{saddest_day}
Schwartz, C. 2020.
\newblock Is Everybody Doing … OK? Let’s Ask Social Media.
\newblock \emph{The New York Times}.
\newblock Accessed: 2021-07-24.

\bibitem[{Schwartz et~al.(2013)Schwartz, Eichstaedt, Kern, Dziurzynski, Lucas,
  Agrawal, Park, Lakshmikanth, Jha, Seligman
  et~al.}]{schwartz2013characterizing}
Schwartz, H.; Eichstaedt, J.; Kern, M.; Dziurzynski, L.; Lucas, R.; Agrawal,
  M.; Park, G.; Lakshmikanth, S.; Jha, S.; Seligman, M.; et~al. 2013.
\newblock Characterizing geographic variation in well-being using tweets.
\newblock In \emph{Proceedings of the International AAAI Conference on Web and
  Social Media}, volume~7, 583--591.

\bibitem[{Schwartz et~al.(2017)Schwartz, Giorgi, Sap, Crutchley, Ungar, and
  Eichstaedt}]{schwartz2017dlatk}
Schwartz, H.~A.; Giorgi, S.; Sap, M.; Crutchley, P.; Ungar, L.; and Eichstaedt,
  J. 2017.
\newblock DLATK: Differential language analysis ToolKit.
\newblock In \emph{Proceedings of the 2017 Conference on Empirical Methods in
  Natural Language Processing: System Demonstrations}, 55--60.

\bibitem[{Syed and Spruit(2017)}]{syed2017full}
Syed, S.; and Spruit, M. 2017.
\newblock Full-text or abstract? examining topic coherence scores using latent
  dirichlet allocation.
\newblock In \emph{2017 IEEE International conference on data science and
  advanced analytics (DSAA)}, 165--174. IEEE.

\bibitem[{Tynes et~al.(2019)Tynes, Willis, Stewart, and
  Hamilton}]{tynes2019race}
Tynes, B.~M.; Willis, H.~A.; Stewart, A.~M.; and Hamilton, M.~W. 2019.
\newblock Race-related traumatic events online and mental health among
  adolescents of color.
\newblock \emph{Journal of Adolescent Health}, 65(3): 371--377.

\bibitem[{Wilkins, Livingstone, and Levine(2019)}]{wilkins2019whose}
Wilkins, D.~J.; Livingstone, A.~G.; and Levine, M. 2019.
\newblock Whose tweets? The rhetorical functions of social media use in
  developing the Black Lives Matter movement.
\newblock \emph{British Journal of Social Psychology}, 58(4): 786--805.

\bibitem[{Williams et~al.(2018)Williams, Metzger, Leins, and
  DeLapp}]{williams2018assessing}
Williams, M.~T.; Metzger, I.~W.; Leins, C.; and DeLapp, C. 2018.
\newblock Assessing racial trauma within a DSM--5 framework: The UConn
  Racial/Ethnic Stress \& Trauma Survey.
\newblock \emph{Practice Innovations}, 3(4): 242.

\bibitem[{Williamson, Trump, and Einstein(2018)}]{williamson2018black}
Williamson, V.; Trump, K.-S.; and Einstein, K.~L. 2018.
\newblock Black lives matter: Evidence that police-caused deaths predict
  protest activity.
\newblock \emph{Perspectives on Politics}, 16(2): 400--415.

\bibitem[{Wu et~al.(2021)Wu, Gallagher, Alshaabi, Adams, Minot, Arnold, Welles,
  Harp, Dodds, and Danforth}]{wu2021say}
Wu, H.~H.; Gallagher, R.~J.; Alshaabi, T.; Adams, J.~L.; Minot, J.~R.; Arnold,
  M.~V.; Welles, B.~F.; Harp, R.; Dodds, P.~S.; and Danforth, C.~M. 2021.
\newblock Say Their Names: Resurgence in the collective attention toward Black
  victims of fatal police violence following the death of George Floyd.
\newblock arXiv:2106.10281.

\bibitem[{Yang(2016)}]{yang2016narrative}
Yang, G. 2016.
\newblock Narrative agency in hashtag activism: The case of \#BlackLivesMatter.
\newblock \emph{Media and communication}, 4(4): 13.

\bibitem[{Zamani et~al.(2020)Zamani, Schwartz, Eichstaedt, Guntuku, Ganesan,
  Clouston, and Giorgi}]{mz-2020-understanding}
Zamani, M.; Schwartz, H.~A.; Eichstaedt, J.; Guntuku, S.~C.; Ganesan, A.~V.;
  Clouston, S.; and Giorgi, S. 2020.
\newblock Understanding Weekly COVID-19 Concerns through Dynamic
  Content-Specific LDA Topic Modeling.
\newblock In \emph{Proceedings of the Third Workshop on Natural Language
  Processing and Computational Social Science}. Association for Computational
  Linguistics.

\end{thebibliography}

\end{document}